\documentclass[runningheads]{llncs}
\usepackage{booktabs}
\usepackage{graphicx}
\usepackage{hyperref}
\usepackage{graphicx}
\usepackage{cite}
\usepackage{amsmath,amssymb,amsfonts}
\usepackage{algorithmic}
\usepackage{graphicx}
\usepackage{textcomp}
\usepackage{xcolor}
\usepackage{graphicx}
\usepackage{tikz}
\usetikzlibrary{matrix}
\usepackage{cite}
\usepackage{amsmath,amssymb,amsfonts}
\usepackage{algorithmic}
\usepackage{textcomp}
\usepackage{hyperref}
\usepackage{booktabs}
\usepackage{xspace}
\usepackage{svg}
\usepackage{amsmath}
\usepackage{footnote}
\usepackage{makecell}
\usepackage{multirow}
\usepackage{enumitem}
\usepackage{array}
\usepackage[nolist,nohyperlinks]{acronym}
\usepackage{hyperref}
\usepackage{subcaption}
\usepackage{caption, subcaption}
\usepackage[acronym]{glossaries}
\usepackage{graphicx}
\usepackage{multirow}
\usepackage{todonotes}
\usepackage{comment}
\usepackage{adjustbox}
\usepackage{amsmath}
\usepackage[utf8]{inputenc}
\usepackage[english]{babel}
\usepackage{color}
\usepackage{amsfonts}
\usepackage{url}
\usepackage[mathscr]{euscript}
 \let\mathscr\relax%
\usepackage[scr]{rsfso}
\usepackage{caption}
\usepackage[labelfont=bf]{caption}
\usepackage{hyperref}
\usepackage{graphicx}
\usepackage{cite}
\usepackage{amsmath,amssymb,amsfonts}
\usepackage{algorithmic}
\usepackage{graphicx}
\usepackage{textcomp}
\usepackage{xcolor}
\usepackage{graphicx}
\usepackage{tikz}
\usetikzlibrary{matrix}
\usepackage{cite}
\usepackage{amsmath,amssymb,amsfonts}
\usepackage{algorithmic}
\usepackage{textcomp}
\usepackage{hyperref}
\usepackage{booktabs}
\usepackage{xspace}
\usepackage{svg}
\usepackage{amsmath}
\usepackage{footnote}
\usepackage{makecell}
\usepackage{multirow}
\usepackage{enumitem}
\usepackage{array}
\usepackage{tabularx,booktabs}
\newcolumntype{C}{>{\centering\arraybackslash}X} 
\usepackage{multicol}
\usepackage{algorithm2e}
\usepackage{booktabs,ragged2e}
\usepackage[utf8]{inputenc}
\usepackage{todonotes}
\usepackage{lineno}
\usepackage{makecell}
\usepackage[flushleft]{threeparttable}
\usepackage{subcaption}
\usepackage{color}
\usepackage[nolist,nohyperlinks]{acronym}
\usepackage{footnote}
\usepackage[perpage]{footmisc}
\usepackage[autostyle]{csquotes}
\makesavenoteenv{tabular}
\acrodef{DL}{Deep learning}
\acrodef{CNN}{Convolutional Neural Network}
\acrodef{ML}{Machine Learning}
\acrodef{GI}{gastrointestinal}
\acrodef{AI}{Artificial Intelligence} 
\acrodef{CADx}{computer aided diagnosis} 
\acrodef{CRC}{colorectal cancer}
\acrodef{DSC}{Dice Coefficient}
\acrodef{mDSC}{Dice Coefficient}
\acrodef{OOD}{Out-Of-Distribution}
\acrodef{SOTA}{State-of-the-art}

\begin{document}

\title{ Classification of Endoscopy and Video Capsule Images using CNN-Transformer Model}

\titlerunning{Classification of Endoscopy and Video Capsule Images using Hybrid Model}

\author{Aliza Subedi\inst{1} \and
Smriti Regmi\inst{1} 
\and
Nisha Regmi\inst{2}
\and
Bhumi Bhusal\inst{2}
\and
Ulas Bagci\inst{2}
\and
Debesh Jha\inst{2}}
\institute{Pashchimanchal Campus, Nepal\\
\and
Machine \& Hybrid Intelligence Lab, Department of Radiology, Northwestern University, Chicago, USA\\
}
%
\maketitle      
\begin{abstract}
Gastrointestinal cancer is the leading cause of cancer-related incidence and death. Therefore, it is important to develop a novel computer-aided diagnosis system for early detection and enhanced treatment. Traditional approaches rely on the expertise of gastroenterologists to identify diseases. However, it is a subjective process, and the interpretation can vary even between expert clinicians. Considering recent progress in classifying gastrointestinal anomalies and landmarks in endoscopic and video capsule endoscopy images, this study proposes a hybrid model incorporating the advantages of Transformers and  \acp{CNN} for enhanced classification performance. Our model utilizes DenseNet201 as a CNN branch to extract local features and integrates the Swin Transformer branch for global feature understanding. Both of their features are combined to perform the classification task. For the GastroVision dataset, our proposed model demonstrates excellent performance with Precision, Recall, F1 score, Accuracy, and Matthews Correlation Coefficient (MCC) of 0.8320, 0.8386, 0.8324, 0.8386, and 0.8191, respectively, showcasing its robustness against class imbalance dataset and surpassing other CNNs as well as Swin Transformer model. Similarly, for the Kvasir-Capsule, a large video capsule endoscopy dataset, our model surpassed all other models, thereby achieving overall Precision, Recall, F1 score, Accuracy, and MCC of 0.7007, 0.7239, 0.6900, 0.7239, and 0.3871. Moreover, we generated saliency maps to explain our model’s focus areas, showing its reliable decision-making process. The results underscore the potential of our hybrid CNN-Transformer model in aiding the early and accurate detection of \ac{GI} anomalies. 

\keywords{Swin Transformer \and Deep learning \and Image classification \and Gastrointestinal tract \and GastroVision \and Kvasir-Capsule}
\end{abstract}

\section{Introduction}
Medical image analysis has become increasingly essential in the diagnosis and prognosis of numerous medical conditions. 
One of the leading causes of cancer death is colorectal cancer~\cite{siegel2024cancer}. It often begins as a growth called a polyp inside the colon or rectum. While not every polyp evolves into cancer, early identification and removal can halt the progression to cancer.
Detecting these conditions early through screening significantly boosts survival chances, since it facilitates intervention during earlier, more treatable stages of the diseases. However, diagnosing these diseases is very time-consuming and tedious. Therefore, diagnostic tools enhanced with \ac{AI} have been developed lately to aid healthcare professionals in more effectively identifying and addressing these health issues.  This approach also improves the quality of medical image analysis and reduces the time and resources needed for proper diagnosis, thereby improving the overall efficiency in healthcare sectors.


The advancement in deep learning algorithms, especially CNNs, which include blocks of convolutional layers, pooling layers, and fully connected layers, has significantly impacted computer vision tasks. Acting as foundational networks, CNNs have demonstrated remarkable effectiveness across various computer vision tasks, such as image classification~\cite{li2014medical}, object detection~\cite{papageorgiou1998general}, and image segmentation~\cite{alom2018recurrent}. CNNs are inherently predisposed to identify patterns in data, providing flexible and adaptive operations that can capture the specific characteristics of the data. They are also effective when used to analyze images obtained from endoscopy~\cite{ahmed2022classification}, leading to more accurate and efficient diagnoses. Recent advancements in deep learning have witnessed the adaptation of transformers, originally designed for natural language processing tasks, to the domain of image analysis. A notable example of this transformative paradigm shift is the Swin Transformer, which demonstrates considerable potential for image classification tasks~\cite{liu2021Swin}. It offers a hierarchical vision by using shifted windows, enabling it to capture both global and local features of images.

This study presents an innovative approach to endoscopic image classification that combines the strengths of both CNNs and Swin Transformers. The proposed model, which uses a DenseNet201~\cite{huang2017densely} as the CNN branch and the Swin Transformer as the other branch, can efficiently classify endoscopic images.  Our extensive experiments demonstrate the model's robustness and superior performance on the two medical GI datasets, paving the way for its application in clinical workflows.
 \vspace{1mm}

The main contribution of our work is as follows: 
\begin{enumerate}
\item We have presented a hybrid architecture for a gastrointestinal (GI) image classification task that effectively combines the capabilities of CNNs and Swin Transformer, bringing a new approach to GI image analysis.

\item We interpreted and visualized the performance of our models using saliency maps.  This approach provides in-depth insights into the model's decision-making process about which parts of the input data are most influential in the model's predictions.


\item Our Hybrid model is able to achieve the MCC of 0.8191 for the GastroVision dataset~\cite{jha2023GastroVision} (highly imbalanced data with 22 classes) and MCC of 0.3871 for the Kvasir-Capsule dataset~\cite{smedsrud2021kvasir} (imbalance data with 11 classes), outperforming the standalone DenseNet201, Swin-T and other CNN methods across all evaluation metrics. 


\end{enumerate}

\section{Related Work}
Our work relates to CNN, Transformer, and GI tract diseases and findings. Here, we thoroughly review the literature that bears significant relevance to these areas.
\vspace{-2mm}

\subsection{Gastrointestinal disease}
Recent studies on the classification of endoscopic images have proposed several deep learning-based techniques~\cite{lopez2023boosting, chou2023preparing, chang2021deep}.
For instance, Gamage et al.~\cite{gamage2019gi} presented the implementation of aggregation of deep learning features to predict anomalies related to digestive tract diseases. The ensemble involves the use of pre-trained CNN models such as DenseNet-201~\cite{huang2017densely}, ResNet-18~\cite{he2016identity}, and VGG-16~\cite{simonyan2014very} as the core feature extractors, integrated with a Global Average Pooling (GAP) layer. Similarly, Thambawita et al.~\cite{thambawita2018medico} presented five different methods for GI tract disease classification task. Their combination of Densenet-161 and Resnet-152 with an additional MLP showcased highest performance. Afriyie et al.~\cite{afriyie2022gastrointestinal} presented Dn-CapsNets for identifying gastrointestinal tract diseases in the Kvasir-v2 dataset~\cite{Pogorelov:2017:KMI:3083187.3083212}. Srivastava et al.~\cite{srivastava2022video} introduced FocalConvNet, a network that merges focal modulation with lightweight convolutional layers, for the classification of anatomical and luminal findings within the Kvasir-Capsule dataset. 

\vspace{-2mm}

\subsection{Transformer network}
Recently, architectures based on transformers have become popular due to their proficiency in handling long-term dependencies. 
The transformer architecture, which is commonly utilized for natural language processing, has been shown by Dosovitskiy et al.~\cite{dosovitskiy2020image} to be effective when employed for computer vision tasks as well. According to their research, Vision Transformer (ViT) performs exceptionally well on image classification tasks when applied to sequences of image patches. Similarly, Touvron et al.~\cite{touvron2021training} developed a novel teacher-student strategy for training convolution free transformers on ImageNet, using a distillation token to facilitate learning from the teacher through attention. Similarly, Usman et al.~\cite{usman2022analyzing} and Matsoukas et al.~\cite{matsoukas2021time} compared the performance of Transformers and CNN in various image classification tasks.
Likewise, Tang et. al~\cite{tang2023transformer} proposed a Transformer-based Multi-task Network to analyze GI-tract lesions automatically by combining the benefits of both the CNN and Transformer.



\begin{figure}[!t]
    \centering
    \includegraphics[trim=0.1cm 13cm 7.68cm 13cm, clip, width=1\textwidth]{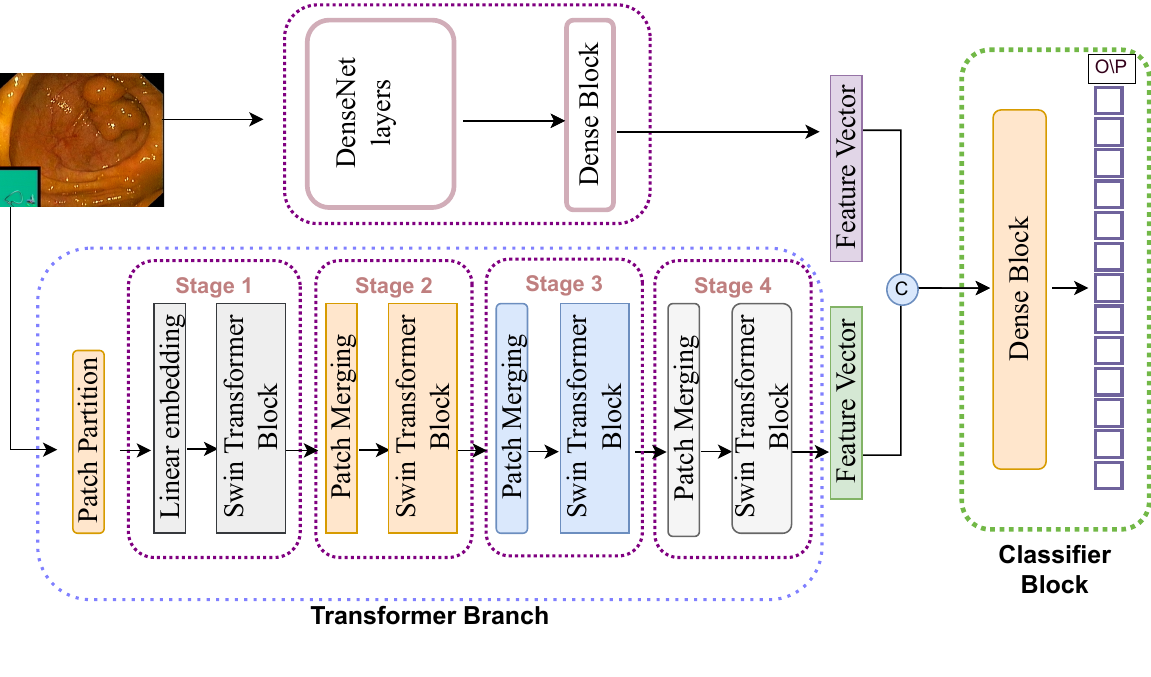}
    \vspace{-7mm}
    \caption{Schematic representation of the Hybrid CNN-Transformer model.}
   \vspace{-5mm}
    \label{fig:block}
\end{figure}

\vspace{-2mm}

\section{Methodology}
\vspace{-2mm}
Here, we merged the capabilities of CNN and transformer networks to address the challenge of identifying the images encountered in endoscopic procedures. Our model has three components - \textit{CNN branch}, \textit{Transformer branch}, and \textit{Classifier block} as shown in Figure~\ref{fig:block}. The CNN branch captures detailed, local features while the Transformer branch focuses on broader, global aspects. These are subsequently merged into a single feature vector. This concatenated feature vector is fed into the \textit{Classifier block} for the final classification task. The dense block in the figure consists of several layers, including a dense layer with 256 units, followed by a LeakyReLU activation with an alpha of 0.1, batch normalization, and a dropout layer with a 0.5 rate.

\subsection{Transformer Branch}

For the transformer branch, we utilized a variant of Swin Transformer, namely Swin-T, where `T' refers to the tiny. The model was initialized with weights pre-trained on the ImageNet1K dataset~\cite{russakovsky2015imagenet}. Here, an input RGB image is first divided into non-overlapping patches of $4\times 4$, each patch serving as a token. These patches have a feature dimension of $4 \times 4 \times 3 = 48$. Each patch is then linearly transformed into an embedding. The Swin Transformer model~\cite{liu2021Swin} consists of multiple layers of transformer encoders. Notably, the Swin Transformer replaces the traditional multi-head self-attention module in a Vision Transformer block with a shifting window-based self-attention. This design allows the model to attend to nearby patches while focusing on local context, without needing large attention windows. After processing the embeddings through several transformer layers, the information from various patches is aggregated to produce a single feature vector.

\subsection{CNN Branch}
In this branch, a CNN is used, specifically DenseNet201. We fine-tuned all the layers of the DenseNet201 on our dataset to enable it to extract features pertinent to our specific classification task. These densenet layers are then followed by \textit{Dense block}. 

\begin{figure}[!t]

\scriptsize
\centering
\includegraphics[width=0.84\linewidth]{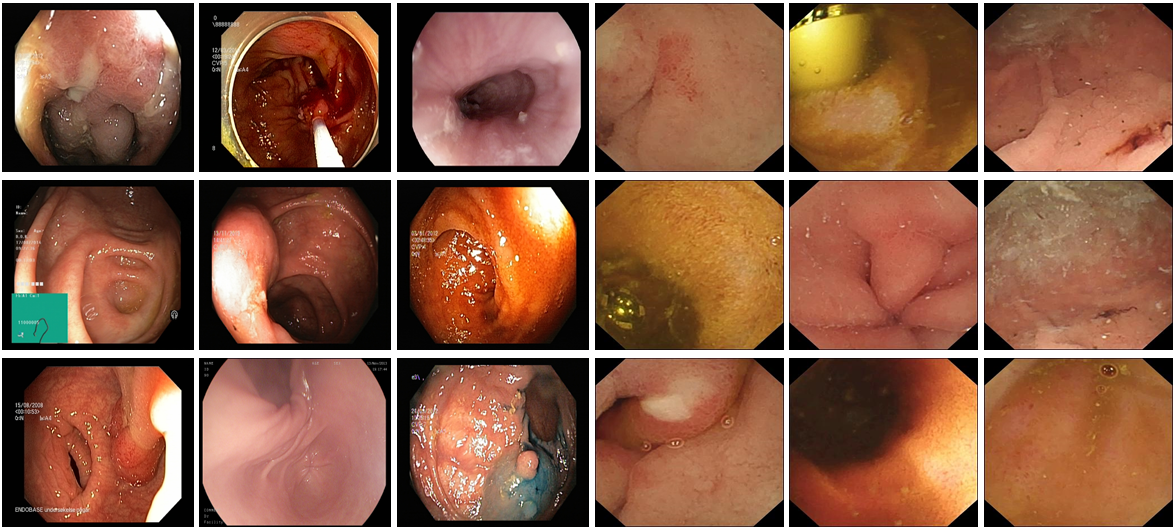}
\caption{Example samples from GastroVision~\cite{jha2023GastroVision} and Kvasir-Capsule~\cite{smedsrud2021kvasir}. The first three columns display images from GI endoscopy (GastroVision dataset), last three columns display from video capsule endoscopy (Kvasir-Capsule).}


\label{fig:datasetsample}
\vspace{-4mm}
\end{figure}

\section{Dataset}

We employed two multi-class publicly available GI endoscopy datasets: GastroVision and the Kvasir-Capsule dataset. Some sample images of the datasets can be observed from the Figure~\ref{fig:datasetsample}.
\begin{enumerate}

\item \textbf{GastroVision~\cite{jha2023GastroVision}: } 
The GastroVision dataset encompasses a wide variety of classes, including anatomical landmarks, pathological findings, cases of polyp removal, and normal or regular findings. A total of 7,930 images from 22 classes were used in the experiment following the dataset provider.

\item \textbf{Kvasir-Capsule~\cite{smedsrud2021kvasir}: } 
Kvasir-Capsule is the largest publicly available video capsule endoscopy dataset, which comprises 44,228 meticulously labeled images representing 13 distinct classes of anatomical and luminal findings. We have utilized 11 out of 13 classes, as some classes have very few samples. 



\end{enumerate}

\section{Experiments}

\subsection{Implementation details}

\subsubsection{Pre-Training :}

In this study, all the models were implemented using Tensorflow\cite{abadi2016tensorflow} framework. We then resized all images to a standard dimension of $224\times 224$ pixels and performed normalization and different data augmentation techniques to mitigate the data limitations. As part of the data augmentation process, we applied several techniques from TensorFlow's Keras utilities, including rescaling pixel intensities, inducing shear transformations, applying rotations of up to 30 degrees, and performing vertical flips. Additionally, we adjusted the brightness to account for variations in lighting conditions. These strategic preprocessing and augmentation measures diversified our dataset and enhanced the generalization capabilities of our implemented architectures. 

\subsubsection{Experiment setup and configuration:} Our experiments revolve around the task of classifying images using various architectures. To optimize our model, we used a trial-and-error approach to adjust various hyperparameters, such as learning rate, batch size, loss function, and optimizer. Specifically, the initial learning rate was set to $1e^{-2}$ for the GastroVision dataset and $1e^{-3}$ for the Kvasir-Capsule dataset, with dynamic adjustments made using the ReduceLROnPlateau scheduler to optimize convergence. We utilized categorical cross-entropy as the loss function, which is suitable for multi-class classification tasks. These hyperparameters were iteratively refined through multiple experimental runs to achieve the best performance metrics for each dataset. The GastroVision dataset is divided into an 80:20 ratio for training and testing, respectively. Similarly, we used~\href{https://github.com/simula/kvasir-capsule/tree/master/official_splits}{official split} (split1) for Kvasir-Capsule dataset.


\subsection{Evaluation metrics}


We used various standard computer vision metrics to evaluate the performance of the models. These include MCC, weighted average precision, F1-score, recall, and overall accuracy. Of all these metrics, we prioritized MCC as the key metric since it is a more reliable statistical measure that yields a high score only when the classifier accurately predicts the majority of positive and negative data instances, and when both positive and negative predictions are mostly correct. We calculated these metrics over stratified 5-fold cross-validation and presented their averages along with their standard deviations. Additionally, we plotted saliency maps to visualize and interpret the performance of our model.

\begin{table*}[!t]
\scriptsize
\centering
\caption{
Quantitative results on the GastroVision~\cite{jha2023GastroVision} dataset. }
\begin{adjustbox}{max width=\textwidth}
\begin{tabular}{l|l|l|l|l|l}
\toprule
\textbf{Method} & \textbf{Precision} & \textbf{Recall} & \textbf{F1-score} & \textbf{ Accuracy} & \textbf{MCC} \\
\midrule

 MobileNetV2~\cite{sandler2018mobilenetv2} &  0.7308 $\pm$ 0.0356 & 0.7400 $\pm$  0.0306 & 0.7318 $\pm$ 0.0332 & 0.7400 $\pm$ 0.0306 & 0.7083 $\pm$ 0.0347 \\
\hline

ResNet50~\cite{he2016deep}  &  0.7151 $\pm$ 0.0154& 0.7320 $\pm$ 0.0159 & 0.7170  $\pm$ 0.0179 & 0.7320 $\pm$ 0.0159 & 0.6988 $\pm$ 0.0182 \\
\hline

Xception~\cite{chollet2017xception} &  0.7410 $\pm$ 0.0050 & 0.7499 $\pm$ 0.0032 & 0.7430 $\pm$  0.0040 & 0.7499 $\pm$ 0.0032 & 0.7195 $\pm$ 0.0036 \\
\hline

InceptionV3~\cite{szegedy2016rethinking} &  0.7756 $\pm$ 0.0070 & 0.7847 $\pm$ 0.0049 & 0.7774 $\pm$ 0.0053 & 0.7860 $\pm$ 0.0048 & 0.7600 $\pm$ 0.0054 \\
\hline

Densenet201~\cite{huang2017densely} &  0.8056 $\pm$ 0.0062 & 0.8112 $\pm$ 0.0052 & 0.8046  $\pm$ 0.0056 & 0.8112 $\pm$ 0.0052 & 0.7886 $\pm$ 0.0059 \\
\hline
Swin-T~\cite{liu2021Swin} & 0.8075 $\pm$ 0.0023 & 0.8148 $\pm$ 0.0038 & 0.8082 $\pm$ 0.0031 & 0.8148 $\pm$ 0.0038 & 0.7924 $\pm$ 0.0042  \\
\hline
\shortstack{\textbf{DenseNet201~\cite{huang2017densely}}\\ \textbf{+ Swin-T~\cite{liu2021Swin}}} & \textbf{0.8320} $\pm$ 0.0204 & \textbf{0.8386} $\pm$ 0.0221 & \textbf{0.8324} $\pm$ 0.0250 & \textbf{0.8386} $\pm$ 0.0035 & \textbf{0.8191} $\pm$ 0.0038 \\
\hline
\end{tabular}
\end{adjustbox}
\label{table: Result1}

\vspace{-3mm}

\end{table*}

\section{Results}
We examined the trained model's effectiveness through multiple quantitative measures across a 5-fold cross-validation and  the result is 
presented in Table \ref{table: Result1} and Table \ref{table: Result2}. 
Our hybrid CNN-Transformer model's performance is compared against the standalone DenseNet201, Swin Transformer model and other CNN methods. Here, each entry in the table is expressed as the mean $\pm$ SD of the respective performance metric, calculated over the 5-folds of cross-validation.  We have reported the standard deviation (SD) to show the consistency of the model performance across five different folds, where a lower SD indicates consistent performance.

The classification performance of the models trained on GastroVision dataset is shown in
Table~\ref{table: Result1}.
For this dataset, our proposed hybrid model, the combined Swin Transformer and DenseNet201, consistently outperformed the other models across all performance metrics. With a F1 score, accuracy, and MCC of 0.8324, 0.8386 and 0.8191 respectively, our model demonstrated the highest performance. Moreover, the table reveals that MCC of the hybrid model surpasses pretrained CNN methods by over 3\% and exceeds Swin T by 2\%. Most competitive to the Hybrid model is Swin T, with MCC of 0.7924. Moreover, the classification performance of various models trained on the Kvasir-Capsule dataset is presented in Table~\ref{table: Result2}. The hybrid model achieved F1-score, Accuracy, and MCC of 0.6900, 0.7239, and 0.3871, respectively, outperforming all CNN-based methods and the standalone Swin Transformer model. Furthermore, as depicted in table, the MCC of the hybrid model trained on Kvasir-Capsule dataset outperforms pretrained CNN methods by more than 3\% and surpasses Swin T by 2\%. Swin T emerges as the closest competitor to the hybrid model. The disparity in the classification performance in two datasets can be attributed to the characteristics and quality of the datasets. The GastroVision dataset, with 22 classes, likely benefits from higher quality images and more distinct class features. In contrast, the Kvasir-Capsule dataset, despite its larger size, has significant class imbalance which poses a challenge for effective model training and performance.



\begin{table*}[!t]
\scriptsize
\centering
\caption{
Quantitative results on the Kvasir-Capsule~\cite{smedsrud2021kvasir} dataset.}
\begin{adjustbox}{max width=\textwidth}
\begin{tabular}{l|l|l|l|l|l}
\toprule
\textbf{Method} & \textbf{Precision} & \textbf{Recall} & \textbf{F1-score} & \textbf{ Accuracy} & \textbf{MCC} \\
\midrule
 MobileNetV2~\cite{sandler2018mobilenetv2} &  0.6100 $\pm$ 0.0047 & 0.6681 $\pm$ 0.0029 & 0.6283  $\pm$ 0.0028 & 0.6681 $\pm$ 0.0029 & 0.2489 $\pm$ 0.0074 \\
\hline
ResNet50~\cite{he2016deep}  &  0.6257 $\pm$ 0.0020 & 0.6810 $\pm$ 0.0026 & 0.6402  $\pm$ 0.0020 & 0.6810 $\pm$  0.0026 & 0.2785 $\pm$ 0.0055 \\
\hline

Xception~\cite{chollet2017xception}  & 0.5972 $\pm$ 0.0026 & 0.6857 $\pm$ 0.0012 & 0.6173 $\pm$  0.0014 & 0.6857 $\pm$ 0.0012 & 0.2186 $\pm$ 0.0039 \\
\hline

InceptionV3~\cite{szegedy2016rethinking} &  0.5964 $\pm$ 0.0040 & 0.6861 $\pm$ 0.0019 & 0.6207  $\pm$ 0.0027 &  0.6861 $\pm$ 0.0019 & 0.2258 $\pm$ 0.0060 \\
\hline

Densenet201~\cite{huang2017densely} &  0.6726 $\pm$ 0.0041 & 0.7043 $\pm$ 0.0044 & 0.6711  $\pm$ 0.0035 & 0.7043 $\pm$ 0.0044 & 0.3508 $\pm$ 0.0067 \\
\hline
Swin-T~\cite{liu2021Swin} & 0.6951 $\pm$ 0.0081 & 0.7088 $\pm$ 0.0068 & 0.6785 $\pm$ 0.0082 & 0.7088 $\pm$ 0.0068 & 0.3600 $\pm$ 0.0173  \\
\hline
\shortstack{\textbf{DenseNet201~\cite{huang2017densely}}\\ \textbf{+ Swin-T~\cite{liu2021Swin}}} & \textbf{0.7007} $\pm$ 0.0238 & \textbf{0.7239} $\pm$ 0.0105 & \textbf{0.6900} $\pm$ 0.0180 & \textbf{0.7239} $\pm$ 0.0105 & \textbf{0.3871} $\pm$ 0.0333 \\
\hline
\end{tabular}
\end{adjustbox}
\label{table: Result2}
\vspace{-4mm}
\end{table*}


\section{Discussion}


\subsection{Limitations and open challenges}
We encountered several problems, including data limitations and problems with resource usage during the study. Moreover, the inherent imbalance within the datasets added complexity to the training process, requiring careful handling to prevent biased model outcomes. We tried solving the problem of data limitations using various data augmentation techniques.  To date, we have not yet engaged with clinicians to incorporate their invaluable expertise. Recognizing their critical role, we see a significant opportunity to collaborate with them to further refine our model's architecture and enhance its clinical relevance. Moving forward, our aim is to foster interdisciplinary collaboration, paving the way for the development of more effective and clinically applicable AI solutions.



\begin{figure*}[!h]
\centering
\includegraphics[trim=0.18cm 0.1cm 0.5cm 0.1cm, clip, width=0.85\linewidth]{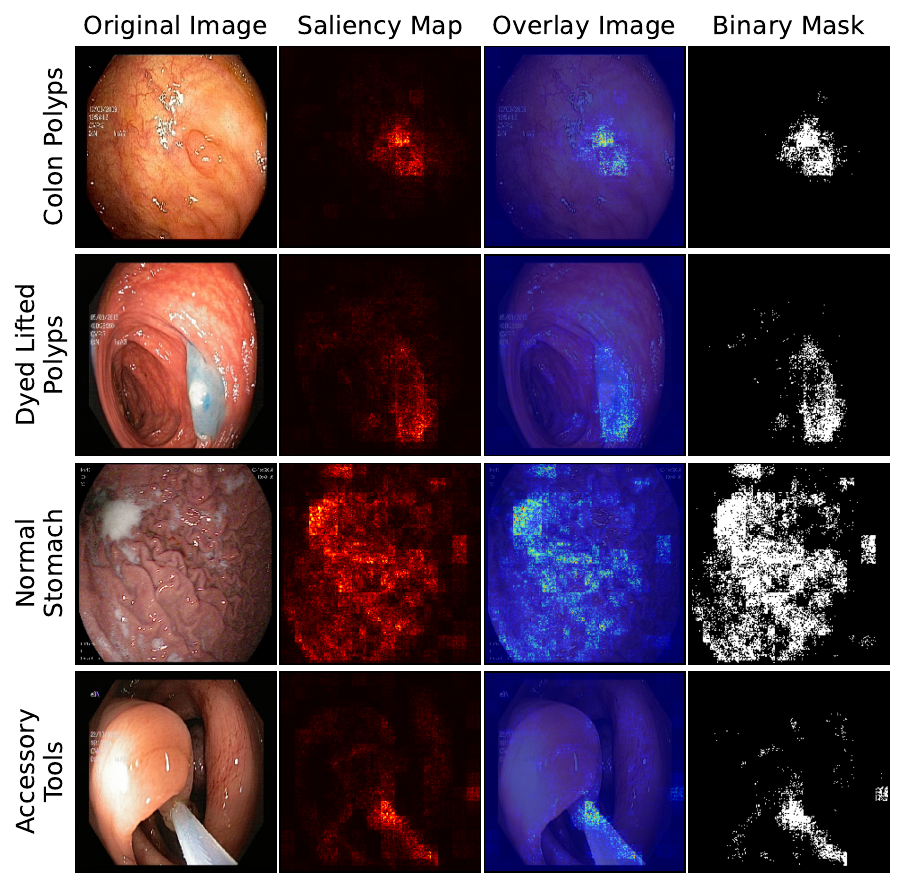} 
\vspace{-1mm}

\caption{Saliency maps visualization for GastroVision dataset for four representative classes. The Figure shows the significant regions in the input image that contribute to the model's decision. } 
\label{fig:saliency1}

\end{figure*}

\subsection{Visualization using saliency maps}

 
A saliency map is a popular technique for visualizing the areas in the image that the model prioritized as the most important when making a prediction. To understand and interpret the behavior of our model, we have generated saliency maps for different representative classes, as shown in Figure~\ref{fig:saliency1}. These saliency maps are produced by computing the gradients of the model's output with respect to its input image, highlighting the pixels that have the most significant impact on the model's decision-making process.

In the figure, each class consists of four panels: the original image, the saliency map highlighting the regions with the highest influence on the model's decision, the overlay of the saliency map on the original image, and a binary mask derived from the saliency map. This comprehensive visualization helps in interpreting the model’s focus areas, further strengthening its reliability and trustworthiness, which are essential in high-risk medical applications, such as GI disease classification tasks. For instance, in the \textit{``Colon Polyps"} example, the saliency map emphasizes the area where the polyp is located, indicating that the model has learned to focus on the key feature relevant to diagnosing polyps. Similarly, for the \textit{``Dyed Lifted Polyps"} class, the saliency map shows a strong focus on the dyed region, which is crucial for identifying the presence of a lifted polyp. These focused regions in the saliency maps correspond to the medically significant features that are critical for accurate classification, thereby demonstrating the model's interpretability. Additionally, in the \textit{``Accessory Tools"} class, the saliency map highlights the regions where the medical tools are present in the endoscopic images. This level of interpretability is similarly applicable to other classes in the dataset, where the saliency maps consistently highlight the most relevant and significant features required for accurate classification.

By overlaying the saliency map on the original image, as depicted in Figure~\ref{fig:saliency1}, we can see a clear distinction between relevant and irrelevant areas. By providing a visual representation of what the model deems important, saliency maps reveal whether the model is focusing on appropriate features. This is particularly important in medical imaging, where the accuracy and reliability of the model's focus are critical for patient outcomes.

\section{Conclusion}
\vspace{-3mm}
In this work, we proposed a hybrid model for GI image classification that combines the advantages of both CNN and Transformer for improved classification performance. The performance of our proposed models was evaluated using stratified 5-fold cross-validation and compared against individual DenseNet201, Swin Transformer model and other CNN methods. Our experimental results demonstrated that the proposed hybrid model consistently outperformed other methods, achieving an MCC of 0.8191 for the GastroVision dataset and an MCC of 0.3871 for the Kvasir-Capsule dataset. Furthermore, we employed saliency maps to visualize and interpret the decision-making process of our models. Additionally, the success of our model in classifying endoscopic images opens up new possibilities for its application in other medical imaging sectors, potentially enhancing clinical decision-making and improving patient well-being.

\subsubsection{Acknowledgements} This project is supported by NIH funding: R01-CA246704, R01-CA240639, U01-DK127384-02S1, and U01-CA268808.

\newpage
 \bibliographystyle{splncs04}
\bibliography{ref}
\end{document}